# THE SPREADING LAYER OF GX 9+9

Osmi Vilhu[1], Ada Paizis[2], Diana Hannikainen[1], Juho Schultz[1], Volker Beckmann[3],

[1] Observatory Box 14, FIN-00014 University of Helsinki, Finland, E-mail: osmi.vilhu@helsinki.fi
[2] INAF-IASF, Milano, Italy, E-mail: ada@iasf-milano.inaf.it
[3] NASA/GSFC, Greenbelt, USA E-mail: beckmann@milkyway.gsfc.nasa.gov

## ABSTRACT

The spreading layer (SL) of GX 9+9 during the upper banana state was studied using *INTEGRAL* and *RXTE* observations. The SL-area becomes larger with increasing accretion rate while the SL-temperature remains close to the critical Eddington value, confirming predictions of [3] and [7]. However, at low accretion rate the observed temperature is higher and SL-belt shallower than those predicted, requiring confirmation and theoretical explanation.

## 1. INTRODUCTION

GX 9+9 (4U 1728-16) is a bright LMXB atoll-source where the disc penetrates down to the neutron star (NS) surface. The accretion flow is partially dissipated in the inner disc and in the belt-like spreading layer (SL) around the NS equator (see Fig.1). SL is an alternative to the boundary layer (BL) concept. SL-theory was introduced in [3], further developed in [7] and compared to observations in [1] and [7]. The SL-area increases with increasing accretion while the SL-temperature remains practically constant and close to the critical Eddington temperature. Hence, the spectrum basically consists of two black bodies with changing relative luminosities depending on where the accretion is dissipated.

Reference [4], based on *RXTE* and optical observations, and [6], based on *INTEGRAL*/ISGRI observations and including several other LMXB's, are new contributions to the relatively slim literature of GX 9+9.

The orbital X-ray modulation of GX 9+9 has been less than 6 % until January 19, 2005, when it suddenly rose to 18 % [5]. Hence, for the present observations, performed during 2003-2004, the variability was related to accretion, only.

## 2. OBSERVATIONS

The observations consist of *INTEGRAL* core programme scans during March/April 2003, October 2003, February 2004 and simultaneous pointed *RXTE* observations on October 9, 2003. For the average spectrum data from all instruments (JEM-X, IBIS/ISGRI, SPI, PCA and HEXTE) were used, while for the temporal variability in soft X-rays only JEM-X and PCA were used with ~ 2000 -s time resolution.

## 3. MODELING OF THE SL-SPECTRUM

We assume that the disc penetrates down to the NS surface with R = 10 km and its radiation is modeled with the DISKBB-model of the XSPEC11/XANADU software. The spreading layer radiation is modeled with the COMPBB-model. The Comptonization part is needed only above 20 keV below which the spectrum consists essentially of two (changing) black bodies. We keep the disc normalization and Compton optical depth and electron density as temporal constants, after fixing them from the average spectrum (see section 4.1). In this way the accretion changes are reflected through three parameters: the inner disc temperature $T_{in}$, the spreading layer temperature $T_{bb}$ and area $norm_{bb}$. *INTEGRAL* observations were reduced with OSA5.1.

## 4. RESULTS

We study the average spectrum and temporal snapshot spectra separately.

### 4.1 Average Spectrum

The average sum-spectrum was fitted with the DISKBB+COMPBB model between 3 – 100 keV

using data from all *INTEGRAL* and *RXTE* instruments. No photoelectric absorption was needed due to the relatively high Galactic latitude (8 deg). A two per cent systematic error was used resulting in an acceptable fit (see Table 1). Figures 2 and 3 show the fits with instruments and model components separated, respectively.

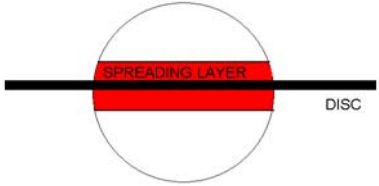

Fig. 1. The spreading layer (SL) on NS surface. The radiation modeled in the present paper comes from the disc + spreading layer.

The electron temperature of the Comptonization part was frozen to 10 keV since only the product $T_e * \tau$ can be meaningfully determined from the spectrum. Further, the electron optical depth itself could be smaller if the inner disc photons were included as seed photons. Assuming an inclination (the tilt between the line of sight and disc normal) of $70°$ results in a distance of 7 kpc. Both of these values can be accepted although they are not known for sure [4].

Table 1. Best-fit parameters of the average spectrum of GX 9+9 (see Figs 2 and 3). In the fit Tin, Tbb, normdisc, normbb and $\tau$ were free parameters. Tin and Tbb are the inner disc and spreading layer temperatures, respectively. Normbb is related to the SL- area. The assumed inner disc radius Rin, inclination incl and distance d are compatible with the disc normalization normdisc. Systematics 0.02 were used.

| *Tin* (keV) | *normdisc* | *Rin* (km) | *incl* (deg) | *d* (kpc) |
|---|---|---|---|---|
| 1.47 +- 0.3 | 68.2 +- 13.0 | 10 (frozen) | 70 +-5 | 7 (frozen) |
| *Tbb* (keV) | *normbb* | *Te* (keV) | *tau* | *redchi2* |
| 1.94 +- 0.4 | 36.2 +- 7.0 | 10 (frozen) | 0.53 +- 0.1 | 1.08 |

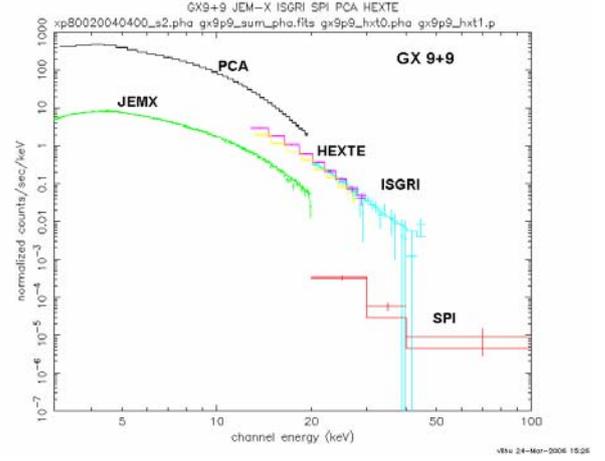

Fig. 2. The average *INTEGRAL* and *RXTE* spectrum of GX 9+9 fitted with the disc + spreading layer radiation model (see Table 1). X-axis: channel energy in keV, Y-axis: normalized counts /keV/s.

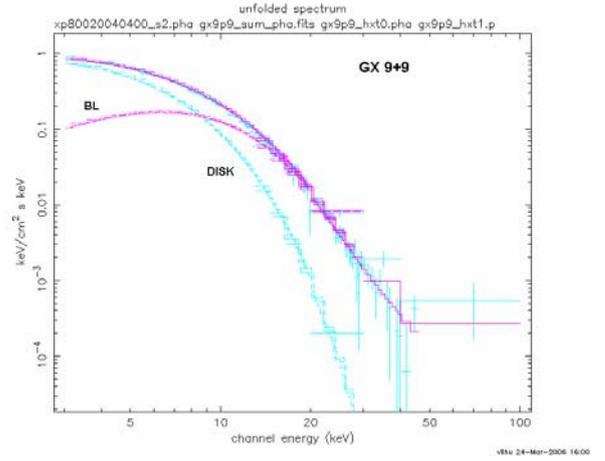

Fig. 3. The spectrum of GX 9+9 separated by the disc and spreading ('boundary') layer (marked as BL) contributions. X-axis: channel energy, Y-axis: keV/cm$^2$/s/keV.

### 4.2 Snapshot Spectra and Comparison with Spreading Layer Models

To study temporal variability we used JEM-X and PCA spectra with around 2000-s exposure times. JEM-X data were limited to less than 3-deg offset angles between GX9+9 and the centre of the field of view, resulting in 49 individual spectra. For *RXTE*/PCA a total of 14 spectra were used.

Each spectrum was fitted with the DISKBB+COMPBB model freezing all parameters of Table 1 except Tin, Tbb and normbb. To have an idea how the behaviour of GX 9+9 during our observations is related to other atoll sources, colors and luminosities from the model spectra were computed as done in [2]. These are shown in Figs. 4 and 5. Apparently GX 9+9 was in the 'upper banana' state during our observations. The colors in Figs. 4 and 5 are corrected for absorption and independent of the instrument response defined as follows: soft color = 4-6.4/3-4 keV,
hard color = 9.7-16/6.4-9.7 keV.

Next we studied how the changes in the accretion rate influence the inner disc temperature and spreading layer temperature and extension. This was possible leaving the inner disc temperature, SL temperature and extension as free parameters and freezing others to the values of Table 1. The results are shown in Figures 6 – 8. The total accretion rate was rather constant (see Fig. 5) and spectral changes were caused mostly by alternation of channelling of the dissipation in the disc and spreading layer (see Fig. 8). The observations confirm the predictions of [3] and [7], except at small SL-accretion, where SL-temperatures are larger than the critical ones and areas are smaller.

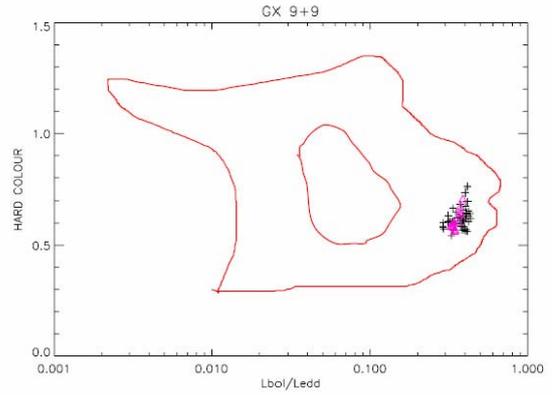

Fig. 5. The same as in Fig. 4 but in the luminosity-color diagram. X-axis: bolometric luminosity in units of the Eddington one, Y-axis: hard color.

However, this may not be statistically significant due to large error bars at low SL-luminosities (see Fig.7). On the other hand, the increase of the observed SL-temperature above the critical one may be real and due to the spectral hardening and color correction at high local SL-luminosities (see [7]).

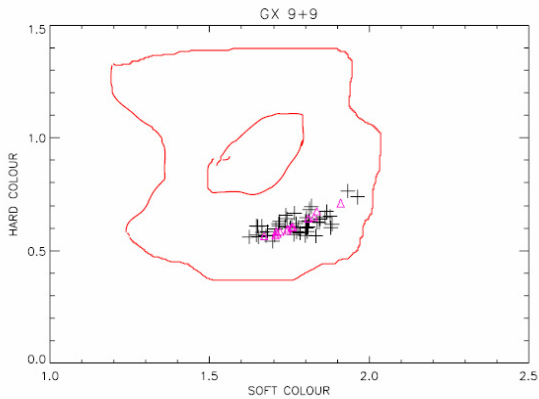

Fig. 4. Snapshot observations of GX 9+9 in the color-color diagram defined in [2]. Plus signs and triangles mark *INTEGRAL* and *RXTE* observations, respectively. The area covered by many atolls in the sample of [2] is marked by hand-drawn lines. X-axis: soft color, Y-axis: hard color.

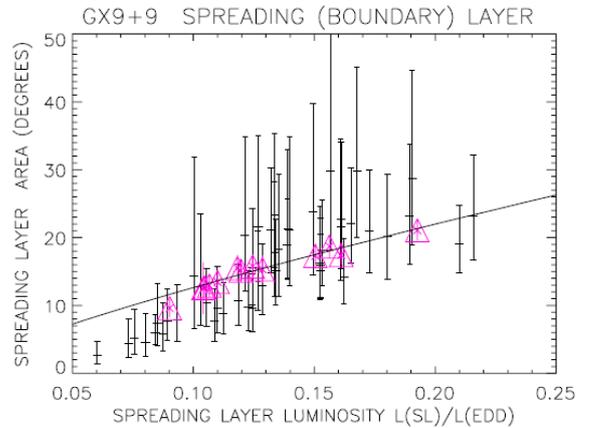

Fig. 6. The spreading layer (SL) latitude belt area (in degrees above and below the NS equator) vs the spreading layer accretion luminosity. The black signs with error bars are from JEM-X pointings, while the triangles are from PCA. The line is the prediction by [3] ( $L^{0.8}$ ). X-axis: SL-luminosity in units of the Eddington one, Y-axis: SL latitude belt area (in degrees).

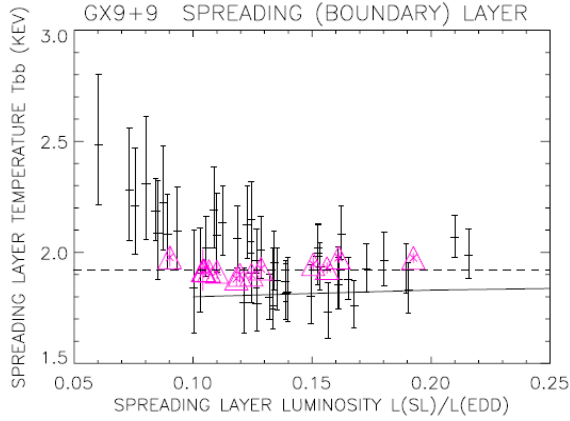

Fig. 7. The same as in Fig. 6 but for the spreading layer temperature vs SL-accretion luminosity. The critical temperature at which the radiation pressure exceeds the NS gravity, is shown by the dashed line ([3]; proportional to NS-gravity$^{1/4}$). The maximum temperature of models in [7] is shown by the solid line. X-axis: The SL-luminosity in units of the Eddington one, Y-axis: SL-temperature in keV

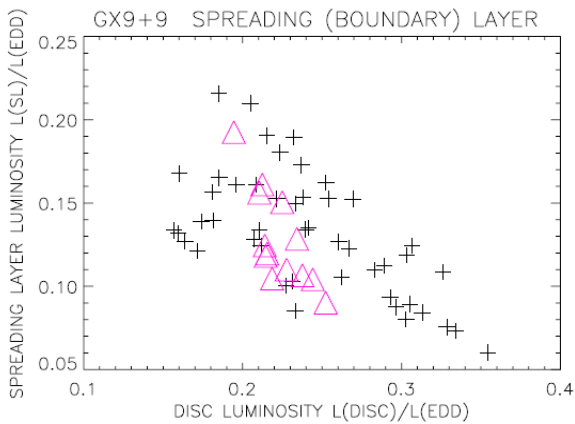

Fig. 8. The spreading layer luminosity vs the disc luminosity as determined by the JEM-X/*INTEGRAL* (plus-signs) and PCA/*RXTE* (triangles) snapshot observations. X-axis: the disc luminosity in units of the Eddington one, Y-axis: the SL-luminosity in units of the Eddington one

## 5. DISCUSSION AND CONCLUSIONS

*INTEGRAL* and *RXTE* spectra of GX 9+9 were modeled with the disc – spreading layer radiation model consisting basically of two black bodies at soft X-rays and a comptonized component above 20 keV. During the observations, the total accretion rate was relatively constant while the major spectral changes were caused by the alternating dissipation in the inner disc and spreading layer (SL) at the NS surface, respectively. The spreading layer temperature was settled close to the critical Eddington temperature. The SL-area was increasing with increasing accretion rate. The results confirm the predictions of [3] and [7]. The observations at low SL-accretion, however, differ from the predictions (observed super-Eddington temperatures). This may be due to the uncertainty of the observations, or due to the color-correction needed at small equatorial belts (diluted Planck [7] like in bursters). High temperatures close to the equator may also be due to the friction between the non-synchronously rotating disc and NS-surface. Observations with high signal-to-noise at low SL-luminosities as well as improved theoretical modeling are required to settle this outstanding question.